\documentclass[prb, nobibnotes,floatfix,superscriptaddress]{revtex4}

\usepackage{amsfonts,amsmath,graphicx,epsfig,latexsym,color,subfigure, afterpage, datetime}

\begin{document}

\title{Electrical tuning of spin current in a boron nitride nanotube quantum dot
 }

\author{Kamal B Dhungana}
\affiliation{Department of Physics, Michigan Technological University, Houghton MI 49931, USA}

\author{Ranjit Pati\footnote{e-mail:patir@mtu.edu}}
\affiliation{Department of Physics, Michigan Technological University, Houghton MI 49931, USA}

\begin{abstract}
\begin{center}
{\bf {Abstract}}
\end{center}

{\footnotesize
Controlling spin current and magnetic exchange coupling by electric field and achieving high 
spin injection efficiency at the same time in a nanostructure coupled to ferromagnetic electrodes have been the outstanding challenges in nanoscale spintronics. A relentless quest is going on
to find new low-dimensional  materials with tunable spin dependent properties to address these challenges. 
Herein, we predict, from first-principles, the transverse-electric-field induced switching in the sign of 
exchange coupling and tunnel magneto-resistance in a boron nitride nanotube quantum dot attached to ferromagnetic nickel contacts. An orbital dependent density
functional theory in conjunction with a  single particle Green's
function approach is used to study the spin dependent current. The origin of switching is attributed to the electric
 field induced modification of magnetic exchange interaction at the interface caused by the Stark effect. 
 In addition, spin injection efficiency is found to vary from 61\% to 89\% depending upon the 
 magnetic configurations at the electrodes.  
 These novel findings are expected to open up a new pathway toward the application of boron 
 nitride nanotube quantum dot in next generation nanoscale spintronics.}
\end{abstract}

\maketitle

\pagebreak

{\bf {I. Introduction}}

Spintronics, which relies on the spin state of the electron to store, transport, 
and process information, has been the subject of intense research since the 
discovery of giant magneto-resistance. \cite{Fert}
With the revolutionary progress in nanotechnology in recent years enabling 
the manipulation of electron spins in  nanoscale tunnel junctions, \cite{ Ouyang, Schmaus, Hauptmann, Sahoo, Urdampilleta, Barraud, Yamada, Horiguchi}  it has 
crossed the boundary of conventional, all metallic, solid state multi-layered structures \cite{Baibich, Binasch, Parkin}
to reach a new frontier, where nano-structures are being used as  controlled spin-carriers. 
When a quantum-confined nanostructure (QCNS) having a non-magnetic character is 
used as a tunnel barrier between two magnetic electrodes, it offers new opportunities 
for the spin manipulation via external electric field {\textemdash} {\it an important prerequisite for nanoscale 
spintronics}; \cite{Dediu, Kim, Sanvito1}  the QCNS in contact with  a ferromagnetic lead loses its non-magnetic property 
due to  the magnetic proximity effect and becomes spin-polarized. \cite{Liu, Atodiresei} The external electric field then not 
only modulates the shape of the spin orbital \cite{Kim} and position of the discrete spin-polarized 
eigen-channels of the QCNS due to the Stark effect,  but also it modifies the electronic and magnetic structure at 
the interface, \cite{Barraud, Dediu, Sanvito1, Waldron, Sanvito2, Ruden}  which plays a dominant role in dictating the spin current behavior of the device.

For example, using  molecular quantum dot,
researchers have already demonstrated giant magnetoresistance effect in  molecular  tunnel (MT) devices. 
\cite{Schmaus, Yamada, Horiguchi}
However, the strong sensitivity of magnetoresistance to the junction structure \cite{Mandal} and the difficulty in achieving 
atomic level controls at the interface  make the implementation of the MT device an arduous task. Other promising nano-structures 
being investigated as spin-carriers are carbon nanotube quantum dots (CNTQD) coupled to ferromagnetic 
electrodes. \cite{Hauptmann, Sahoo, Tsukagoshi} The gate field induced switching of the 
exchange field has already been established in this system. \cite{Hauptmann}
 In addition, the sign modulations of tunnel magnetoresistance (TMR) in both two and three terminal CNTQD-magnetic 
junctions have been reported.  \cite{Sahoo, Wang}
But the difficulty in separating the metallic CNT from the semiconducting one 
poses a significant hurdle toward their practical applications in spintronics. On the other hand, a robust 
semiconducting boron nitride nanotube (BNT), \cite{ Bando, Chopra}  which is structurally similar to CNT, would be an ideal 
tunnel barrier for the spin transport  because its electronic property is independent of its chirality. BNT is also found 
to exhibit a giant response to the transverse electric field \cite{Ishigami} due to strong Stark effect arising from the ionic nature of 
BN bonds {\textemdash} {\it offering new opportunities  to control 
spin current via electric field}. However, 
up until now, no efforts have been made to understand the spin current in such a device. 

Here, we use a boron nitride nanotube quantum dot (BNTQD) 
as a tunnel barrier between two ferromagnetic nickel electrodes  to probe 
the electric field manipulation of spin current. Our first-principles investigation 
reveals transverse electric field ($\varepsilon_g$) induced switching in the sign of  exchange coupling ({\bf J}) and 
tunnel magneto resistance together with a very high spin injection efficiency. The precise role of BNTQD/Ni interface on switching the sign of {\bf J} and TMR 
is identified. In addition, we have observed an intriguing bias dependent switching in spin-polarized current
with a robust negative differential resistance (NDR) feature at a higher ${\varepsilon_g}$. The origin of this 
novel switching property is attributed to the strong field-induced modification of the spin orbitals due to the Stark effect.

The rest of the paper is organized as follows. In the section II, 
we present briefly the computational procedure. Results and
discussions are described in the section III followed by a brief conclusion in the section IV.

{\bf {II. Computational Method}}

The spin-up and  the spin-down components  in the presence of $\varepsilon_g$ are 
calculated within the multichannel Landauer-B$\ddot{u}$ttiker formalism:
$I_{sd}^{\sigma}= \frac{e}{h} \int_{\mu_1}^{\mu_2} T^{\sigma}(E,V_{sd}, \varepsilon_g) {\times} [f(E, \mu_2)-f(E,\mu_1)] {\times} dE$.
$T^{\sigma}(E,V_{sd},\varepsilon_g)$ is the  transmission function obtained from 
the bias dependent spin-polarized Green's function, which is calculated as:
$G^{\sigma}(E, V_{sd}, \varepsilon_g) = [ E   \times  S_{QD} - H_{QD}^{\sigma}(V_{sd}, \varepsilon_g) - \Sigma_{l}^{\sigma}(V_{sd}, \varepsilon_g) - \Sigma^{\sigma}_{r}(V_{sd}, \varepsilon_g)]^{-1} $. $\mu_{1,2}$ are the electro-chemical potentials at the leads, 
which are determined self-consistently (see Ref. 21 for details). 
$H_{QD}(V_{sd}, \varepsilon_g)$ is the bias dependent Kohn-Sham Hamiltonian for the 
BNTQD; $\Sigma^{\sigma}_{l,r}(V_{sd}, \varepsilon_g)$ are the bias dependent spin-polarized self-energy functions, which permit
the BNTQD to exchange its spin-polarized electrons  and energy with  the semi-infinite electrodes. 
We have considered  a chemically bonded junction where the ground state
based DFT has been found to be a good approximation. \cite{Taylor, Brandbyge, Xue, Su, Solomon, Lang} 
The interfacial distance between BNTQD and nickel surface is 
1.9 \AA, which is obtained by minimizing the repulsive interaction within the spin unrestricted density functional theory. 
The energy versus interface distance curve is found to be parabolic around 1.9 \AA$ $ that further justifies  
the use of ground state based DFT in our calculations.
An orbital 
dependent B3LYP hybrid functional for exchange-correlation and an all-electron 6-311g* 
Gaussian basis set \cite{Gaussian} is used to describe the atoms in the device. 
A true dynamically corrected spin-polarized exchange correlation potential \cite{Runge, Sai}
would better represent the transport properties; however, it is difficult to implement  in such a system. It should be noted that the use
of all electron basis set leads to a spin-polarized Hamiltonian matrix of the active scattering region with a dimension  of 1572 $\times$ 1572 for each 
applied bias point. An energy grid of 0.001 eV is used for integration of the transmission. 
The real space approach  adopted here allows us to include 
the most crucial electronic and magnetic structure details of the BNTQD junction from the first-principles. 
The modification in the electronic and magnetic structure of the device due 
to the transverse electric field $(\varepsilon_g)$ is incorporated through the inclusion of 
a dipole interaction term $(\vec{\varepsilon_g} . {\pmb \sum_{i}}\vec{r}(i))$ in the Kohn-Sham Hamiltonian as a perturbation;  the strength of the dipole
interaction is much weaker compared to the electronic interaction. Since the perturbed 
dipole interaction term contains only single particle interactions, we add it 
to the core Hamiltonian during the self-consistent electronic structure 
calculation to include both the first and higher order Stark effects.

{\bf {III. Results and Discussion}}

First, we consider a  prototypical BNTQD-magnetic tunnel junction as shown in  Fig. 1.
For a practical reason, an optimized (6,0) BNT of length 12.3 \AA $ $ is sandwiched 
between two  Ni (111) electrodes to build the open device structure. It should be noted that 
tunnel junction with 
CNTQD channel of diameter of $\sim$ 1 nm  has  been fabricated. \cite {Urdampilleta}
Furthermore, gate field induced amplification in a molecular transistor
with channel length as small as ours' ( $\sim$ 1 nm) has been demonstrated. \cite {Song} 
Since the electrons in the BNT considered here are strongly confined in all three 
dimensions, and there is a lattice mismatch between BNT 
and Ni at the interface, we term it as a BNTQD.
Then, we recourse to a  bias dependent, single particle many-body Green's function approach \cite{Herrmann, Waldron,  Datta, Ventra} to obtain the spin 
dependent current in the BNTQD tunnel device for the parallel spin configuration (PC) and 
the anti-parallel spin configuration (APC) between the electrodes.
Unlike our early work, \cite{Pati} here the bias effect is included
self-consistently. \cite{Mandal} It is important to note that  the self-consistent inclusion of bias
allows us to create an imbalance in carrier concentration at the leads;
on one lead there is a charge surplus and at the other lead there is a charge
depletion resulting to  residual resistivity dipoles. \cite{ Ventra} This is reflected in the bias dependent
 planar average electrostatic potential profiles for the PC and APC (Fig. 2). Both the 
 profiles show an almost linear drop in the potential across the junction with constant potentials at the leads.
 The magnitude of the potential drops at two leads for both  PC and APC  are different, 
 confirming the asymmetric nature of the BNTQD-Ni interfaces at the electronic level.

 {\bf {Spin polarized current.}} Since the spin coherence length is expected to 
be longer than the length of the BNTQD channel considered here, we have adopted a coherent 
spin conserved tunneling approach \cite{Waldron, Rocha, Pati, Mandal,  Ning, Zwolak, Dalgleish}  where the total current is obtained simply by adding 
the spin-up and  the spin-down currents.
The results for the total spin-polarized current  as a function of $\varepsilon_g$ for the PC 
and the APC are presented in Figs. 1a and 1b, respectively. The lead is assumed to have a single 
magnetic domain as shown in the inset of Figs. 1a and 1b. It is noteworthy to mention that in both
cases the spin-down states are found to contribute significantly to the total current ( see Supporting Information). In the absence of $\varepsilon_g$, 
the current for the PC $(I_{PC})$ is found to be higher than the current for the APC $(I_{APC})$. 
A steady increase in current is noted for the bias up to $\sim$0.5 V beyond which a non-linear 
feature in the current is observed for both PC and APC. Remarkably, within the linear current regime, 
with the increase of $\varepsilon_g$, the $I_{PC}$ is found to decrease in contrast to the increase in $I_{APC}$.  
A closer inspection shows a much stronger response to $\varepsilon_g$ in the APC compared to that in the PC. 
For example, at a small bias of 0.2 V, there is  a 16$\%$  decrease in the  $I_{PC}$ compared to an increase
of $221 \%$ in the $I_{APC}$ when $\varepsilon_g$ increases from 0 to 2.04 V/\AA.  
For a higher $\varepsilon_g$, the total currents for both  PC and APC rise initially to 
reach  peak values with the increase of $V_{sd}$ and then drop to  valley points with the subsequent 
increase in $V_{sd}$ before increasing again, revealing  clear NDR features. For $ \varepsilon_g $ = 2.04 V/\AA, 
the peak  to valley current ratio $(I_p/I_v) $ in the APC is found to be 1.7; for  PC the $I_p/I_v$ is 1.4.

{\bf {Tunnel magnetoresistance.}} To quantify this surprisingly contrasting response between the PC and  the APC to $\varepsilon_g$, 
we have calculated the TMR as $(I_{PC} - I_{APC})/I_{APC}$.  Fig. 3 summarizes  the bias dependent 
TMR data for $ \varepsilon_g $ = 0.00 V/\AA$ $ and  2.04 V/\AA. In the absence of the transverse electric field, 
the signs of TMR values
are found to be positive for all the bias points considered here. 
In contrast, for $ \varepsilon_g $ = 2.04 V/\AA, the signs of TMR values
are found to be negative. To elucidate this unique transverse electric field dependent TMR result, we have calculated the TMR as 
a function of $ \varepsilon_g $ at a small bias of 0.2 V (Fig. 1c).
A significant variation in TMR from +23\% to -67\%  with the switching of sign at a critical electric field ($\sim$ 0.8 V/\AA)  is noted. 
We have also performed spin-polarized current calculation using a (7,0) BNTQD 
channel of same length with the same interface distance to check 
whether the switching feature in TMR observed here persists for other diameters. 
Indeed, we have found a similar switching feature in TMR as 
observed for (6,0) BNTQD (See Supporting Information), which confirms  the general nature 
of our observations irrespective of the diameter of the tube. 
It should be noted that a significant diameter dependent band-gap modulation
with  $\varepsilon_g$ has been reported in pristine BNT. \cite{Khoo, Chen} 
The band-gap modulation has been shown  to increase with the tube diameter \cite {Khoo, Chen} and
is found to be  independent of chirality. This clearly suggests that a smaller critical field ($\varepsilon_g^c$) 
than that found in our calculation  would
be suffice to switch the TMR in a BNTQD junction with larger diameter. In addition,  the same order of transverse
 electric field as predicted here has been applied experimentally on BNT of diameter
16.3 $\pm$ 6 {\AA} to observe giant Stark effect, \cite{Ishigami} which implies that our predicted critical field for 
switching TMR would be accessible to the experiment. \cite{Kong, Xu} It is  also worthwhile to note that a similar gate field dependent switching in the sign of 
TMR has been observed at low temperature in a CNTQD-magnetic tunnel junction device. \cite{Sahoo}

{\bf {Magnetic exchange coupling.}} To understand the origin of switching in the sign of TMR, the magnetic exchange coupling, {\bf J} = $E_{PC}-E_{APC}$, 
is calculated as a function of  $\varepsilon_g$ (Fig. 1d). $E_{PC}$ and $E_{APC}$ are total energies  
for the PC and  the APC respectively in the extended system.  At zero $\varepsilon_g$, {\bf J} is found to be positive with the APC being the more 
stable configuration. When we increase  $\varepsilon_g$  from 0 to 2.04 V/\AA,  the value of {\bf J} is found to decrease toward
a negative value with the  switching of  sign  at  $\varepsilon_g$ of $\sim$ 0.8 V/\AA.  
A strong correlation is found between the variation of {\bf J} and TMR; the switching of {\bf J} is found at a slightly higher $\varepsilon_g$ than TMR. 
This can be understood from the fact that {\bf J} calculation does not consider the imaginary part of the Hamiltonian as incorporated in TMR calculation  for the open device.
To gain deeper insights into the cause of sign reversal in {\bf J}, we examine the spin profile of the device as a function of $\varepsilon_g$. 
Due to strong exchange interactions between the electrons  at the interface, the non-magnetic BNTQD becomes spin polarized and the 
atoms that are in close proximity to Ni gain substantial magnetic property; \cite{Liu, Atodiresei} the interface now acts as 
a spin-interface. \cite{Sanvito2, Waldron} Since we have B atoms at the one interface and N atoms at the other, there is 
an asymmetric spin profile at the interfacial atoms.  For example, in the case of APC, the average magnetic moment per atom ($\bar{\mu}$) in the nitrogen
layer at the close vicinity of Ni lead ($\sim$ 1.3 $\mu_B$ per atom) changes from  - 0.12 $\mu_B$ to -0.07 $\mu_B$ when we increase $\varepsilon_g$ from 0 to 2.04 V/\AA;
in the case of boron-nickel interface layer, 
 $\bar{\mu}$ for boron changes from -0.01 $\mu_B$ to -0.03 $\mu_B$. For PC, $\bar{\mu}$ in the boron layer at the interface  decreases from 0.16 $\mu_B$ to 0.13 $\mu_B$
by changing $\varepsilon_g$ from 0 to 2.04 V/\AA; only a small  change from 0.47 $\mu_B$ to 0.48 $\mu_B$ is noted for the $\bar{\mu}$ in the nitrogen layer at the interface.
This spin profile at the interface is shown schematically in the inset of Fig. 1d.
 For $\varepsilon_g$  $<$ $\varepsilon^c_g$, 
the strong negative exchange interaction between Ni and N at the interface for the APC 
(favored by the Hund's rule) explains  the stability of the anti-parallel configuration over the PC. 
For $\varepsilon_g$ $>$ $\varepsilon^c_g$, there is
a substantial decrease in the magnetic moment of the N at the interface for the APC
resulting in a lower negative exchange interaction at the Ni/N spin-interface; at the same time, the magnetic 
moment of the B at the other interface increases leading to a stronger positive exchange 
interaction  at the Ni/B spin-interface. Conversely, for $\varepsilon_g$ $ >$ $\varepsilon^c_g$, a substantial decrease 
in the magnetic moment at the B for the PC leads to a weaker positive exchange interaction 
at Ni/B spin-interface. This makes the PC more stable than the APC for $\varepsilon_g$  $>$ $\varepsilon^c_g$. 
Thus, unambiguously, we confirm that the electric field manipulation of spin-interface 
is the main cause for switching of  {\bf J} and TMR. Now the question arises: What is the mechanism
that causes the change  in spin profile at the interface between PC and APC ?  To answer this, we calculate
electric dipole moment $\alpha^j$ and polarizability $\beta^{jk}$ for PC and APC (shown in the Table I).
Since the y-components of dipole moment and polarizability for PC and APC are distinct, each spin configuration
responds uniquely to the transverse electric field due to Stark effect. \cite{Kim} This, in fact, results in an energy cross
over and switching of {\bf J}. Next, we turn our focus to another important factor, the spin injection co-efficient,   
$\eta$, which dictates the spin injection efficiency from Ni electrode to BNTQD. 
It should be noted that spin injection  into a semiconductor
from a ferromagnetic contact can be measured using spin-resolved two-photon photoemission technique. \cite{Cinchetti}   We have calculated the bias dependent $\eta$ 
as: \cite{Waldron} $\eta$ = $(I_{up}-I_{down})/(I_{up}+I_{down})$; $ I_{up}$ and $I_{down}$
refer  to the spin-up and spin-down components of the current, respectively. For PC, the maximum and minimum spin injection factors ($\eta_{max}$ and
$\eta_{min}$) are found to be -0.89 and -0.81,  respectively (Fig. 4a). In APC, $\eta_{max}$ and $\eta_{min}$ are found 
to be -0.74 and -0.61, respectively ( Fig. 4b). These high values of $\eta$ suggest that the  Ni/BNTQD spin-interface 
acts as an natural spin-selective  tunnel barrier for spin injection.

{\bf { Spin dependent transmission.}} To further our understanding of  the observed $\varepsilon_g$ dependent spin current behavior in  PC and APC, 
we have analyzed the $T^{\sigma}(E, V_{sd}, \varepsilon_g)$. 
For brevity, the results for $ T^{\sigma}$ at two representative 
$\varepsilon_gs$ (0 and 2.04 V/\AA) are summarized in Figs. 5a and 5b.  In both PC and APC, 
we find a substantially higher contribution to the transmission from the spin-down states, 
which explains the higher observed spin-down current (see Supporting Information). A significant broadening occurs 
in the spin-down case for both PC and APC, which can be inferred from the spilling of Ni 
spin-down density of states (SD-DOS) into the BNTQD due to the strong coupling at the interface and a much higher SD-DOS of the Ni-lead 
at the Fermi energy. The asymmetry in $T^{\sigma}$ between spin-up and spin-down states for the APC is 
expected due to intrinsic structural asymmetry at the interface and the bias induced electronic asymmetry.  
Transmission data show a much weaker response to $\varepsilon_g$ in  the case of PC as compared to  the  APC,
which is also  reflected in their respective total spin-polarized currents.  A closer inspection of Figs. 5a and 5b 
reveals that the height of the transmission feature near the Fermi energy decreases for the PC with the increase of $\varepsilon_g$
 (Fig. 5a). In contrast, a substantial increase in the height of the transmission feature is observed 
 for the APC near the Fermi energy with increasing $\varepsilon_g$ (Fig. 5b). This clearly explains 
 the observed decrease in current for the PC compared to an increase in current for the 
 APC at a smaller bias with increasing  $\varepsilon_g$. In the absence of $\varepsilon_g$, 
 the transmission height for the PC in the vicinity of the Fermi energy is higher than that in APC resulting a 
 higher $I_{PC}$ than $I_{APC}$ (Figs. 1a and 1b).

{\bf {Nonlinear spin-polarized current.}} Now the question arises: What is the cause for the strong non-linear NDR behavior in 
the spin-polarized current at a higher $\varepsilon_g$? 
To answer this subtle question, we have looked at the $T^{\sigma}$  at $\varepsilon_g$ = 2.04 V/\AA $ $ for both the PC and  the APC.  
Since the APC shows a much stronger non-liner response at a higher $\varepsilon_g$, we have 
summarized  the results for only the APC  ( Fig. 6a);
only four bias points are considered. The transmission height decreases within the chemical potential window (CPW) 
for the spin-down states as $V_{sd}$ 
increases from 0.35 V to 0.72 V. However, since the width of the CPW  is  
much higher for $V_{sd}$ = 0.72 V and the current is dictated by the area under the transmission curve within the CPW, 
we observe a higher current at $V_{sd}$ = 0.72 V. When we increase the bias further to $V_{sd}$ = 1.51 V, in spite of  the increase in CPW 
width, we see a substantial drop in the height of the transmission within CPW resulting in a significant 
drop in current leading to a NDR feature. A similar drop in transmission with increase in bias leading to NDR feature has been observed in 
molecular junction. \cite{McClain} It should be pointed out that $V_{sd}$ = 0.72 V and 1.51 V 
correspond to the peak-current and valley-current position respectively for the APC. 
When we increase $V_{sd}$ to 1.92 V, the spin-up states start contributing significantly within the CPW leading to an increase in total 
current. The next  question is: Why do we see a significant drop in transmission with the increase in 
$V_{sd}$? We examine the bias dependent spin orbital for the APC and its response to $\varepsilon_g$ to answer this
inquiry.  One of the frontier spin-down orbitals (i.e the highest occupied orbital of the active scattering region, HOMO) 
that contributes to the transmission within the CPW is presented in the inset of Fig. 6a.  
A dramatic transformation in  the shape of the spin orbital is noticeable as  $V_{sd}$ changes from 0.35 V to  1.51 V. 
For $V_{sd}$ = 0.35 V, the electron (spin-down) cloud is distributed at both the interfaces despite
some asymmetry in  distribution between the two interfaces. With the increase of 
$V_{sd}$, the asymmetry in electron density distribution between two interfaces increases; 
at $V_{sd}$ = 1.51 V (valley point),  due to the strong field induced orbital mixing, the electron cloud is distributed only at the Ni/B interface 
resulting in a smaller $T^{\sigma}$, and hence $I^{\sigma}$.

{\bf {Stark effect.}} To understand the non-linear response at higher $\varepsilon_g$ in greater detail, we  then examine the Stark shift 
for both PC and APC   for the extended system (Fig. 6b). The Stark shift is calculated as: $\delta E^n$ = $\epsilon_g^n(V_{sd}) - \epsilon_g(V_{sd}=0)$,
 where $n$ corresponds to different participating spin orbitals and ${\epsilon_g}$ is the energy of the spin orbital in the presence of
 $\varepsilon_g$ and $V_{sd}$. H0 {\textemdash} the energy of the spin-down HOMO at equilibrium {\textemdash} is considered as the reference energy, ${\epsilon_g(V_{sd} = 0)}$.
 Since spin-down states dictate the behavior of the current, we have presented the results only for the spin-down states at $\varepsilon_g$ = 0  and 2.04 V/\AA. 
A strong non-linear Stark shift $({\sum_{j}} \alpha^j \varepsilon_g^j + \frac{1}{2} \sum_{j,k}
\beta^{jk} \varepsilon_g^j \varepsilon_g^k + .........)$ for the frontier orbitals at a higher bias is noticeable at $\varepsilon_g$ = 2.04 V/\AA.  
Each spin orbital responds differently to $\varepsilon_g$ as each orbital has a unique electron density distribution 
with distinctive dipole moment and polarizability. APC is found to exhibit a much stronger response to the transverse electric
field (energy level spacing 
between H0 and H1 decreases significantly)  than the 
PC. This can also be inferred from the calculated dipole moment and polarizability (Table I). In the case of the APC, $\alpha^y$ and $\beta^{zy}$ (components along the transverse field direction) are much stronger than that in PC.

{\bf {IV. Conclusions}} 

We have demonstrated electrical manipulation of quantum spin state of the electron 
in a BNTQD-magnetic tunnel junction device to show 
switching in the sign of exchange coupling and tunnel magnetoresistance. Most importantly, the switching feature in tunnel magnetoresistance 
observed here is found to be independent of the diameter of the BNTQD channel, confirming the general nature of our
prediction. Electric field induced 
Stark effect causing a change in magnetic exchange interaction at the interface is found to be the main mechanism behind the switching in sign.
In addition, we have observed a very high spin injection efficiency from nickel electrode to BNTQD. 
We expect the magnitude of the critical electric field  for switching the
sign of tunnel magnetoresistance to decrease with
the increase in diameter of the BNTQD as revealed from the band-gap modulation study in pristine BNT. 
Since the predicted external electric field  for switching the sign of tunnel magnetoresitance  is within the range accessible to the 
experiment, we expect our findings would open up new initiatives for the application of the BNTQD tunnel device in next 
generation spin based nano-scale electronics.

{\bf Acknowledgements}

This work is supported by NSF through Grant No. 1249504. 
The results reported here were obtained using RAMA and Superior $-$ the high performance computing cluster of the 
Michigan Technological University.

\afterpage{%
\begin{table}
\begin{center}
     \begin{tabular}{| l | l |  l   | l | l | l | l | l | l | l | l | p{5cm} |}
     \hline
     &\multicolumn{3}{l |} {Dipole moment(a.u)} &\multicolumn {6}{l |}{\hspace{0.7 in} Polarizability(a.u)} \\
     \hline
     comp. & $\alpha^x$  & $\alpha^y$ & $\alpha^z$ & $\beta^{xx}$ & $\beta^{yx}$&$\beta^{yy}$&$\beta^{zx}$& $\beta^{zy}$ & $\beta^{zz}$  \\ \hline
     PC & 0.079 & 0.143 & 9.594 &  692.78 & -006.03  & 692.53  & 042.62  &  -000.72 & 5215.20   \\ \hline
     APC & 0.952 & 0.788 & 8.578 & 677.26  &  -006.17  &  675.82  &  -022.72  &122.57  & 5898.05  \\ 
     \hline
     \end{tabular}
  \end{center}
  \caption{{\bf Dipole moment \& Polarizability.} Components of dipole moment ($\alpha$) and polarizability ($\beta$) for  parallel spin configuration (PC) and anti-parallel spin configuration (APC).}
  \label{Table 1}
  \end{table}
\clearpage
}

\afterpage{%
\begin{figure}
\epsfig{figure=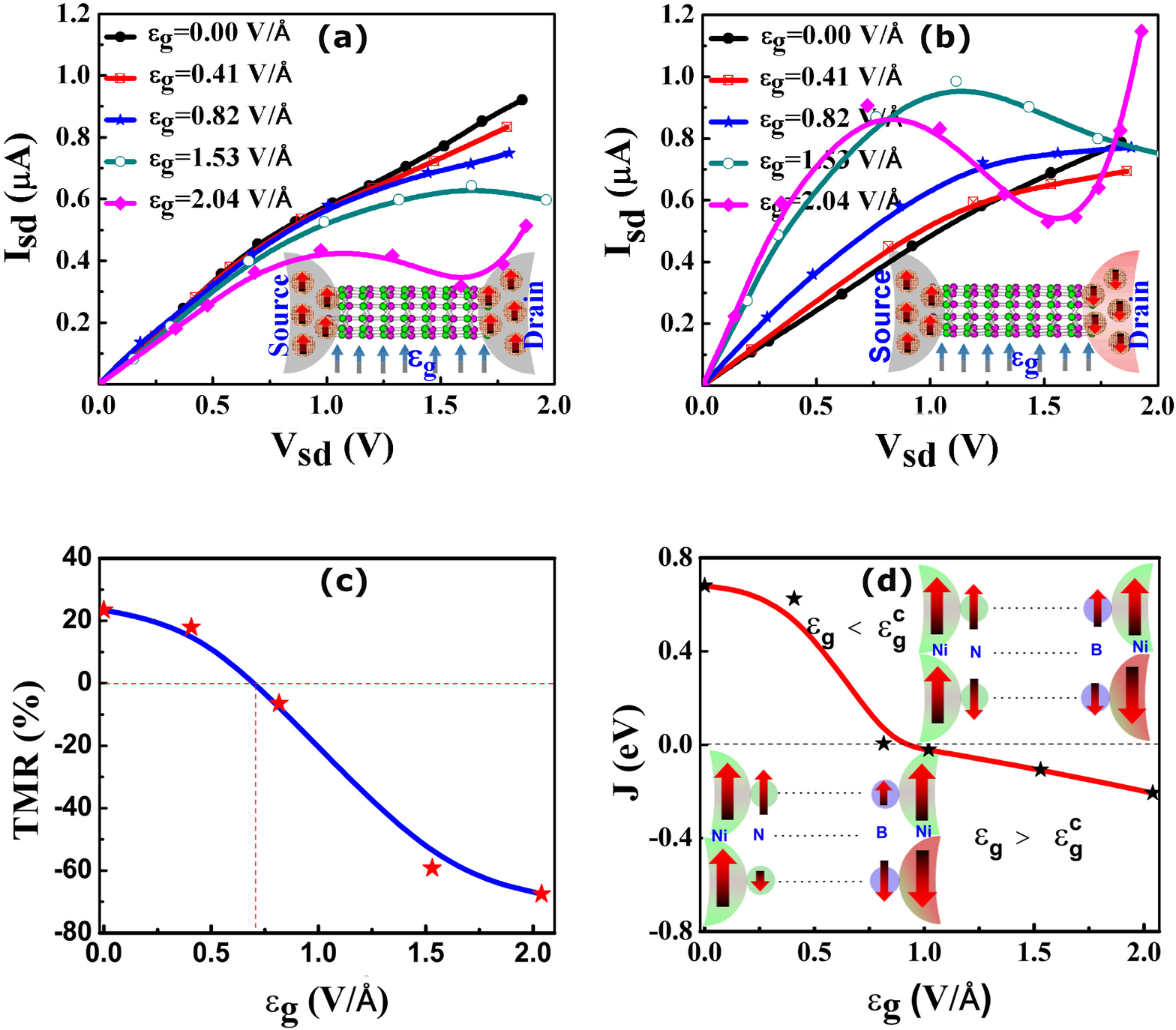, width=440pt}
\caption{{\bf Switching in sign of TMR and J  with transverse electric field.} $I_{sd}$-$V_{sd}$ curves in a BNTQD tunnel junction for (a) PC and (b) APC as a function of
${\varepsilon_g}$. Insets show the schematic junction structures. (c) TMR vs. ${\varepsilon_g}$ at $V_{sd}$ of 0.2 V. (d) 
Exchange coupling ($\bf J$) as function of ${\varepsilon_g}$. Inset shows the ${\varepsilon_g}$ dependent spin-profiles  at the interfaces.
The height and width of the arrow determine the magnitude of magnetic moment. Up and down arrows denote positive and negative 
magnetic moments respectively. }
\end{figure}
\clearpage
}

\afterpage{%
\begin{figure}
\epsfig{figure=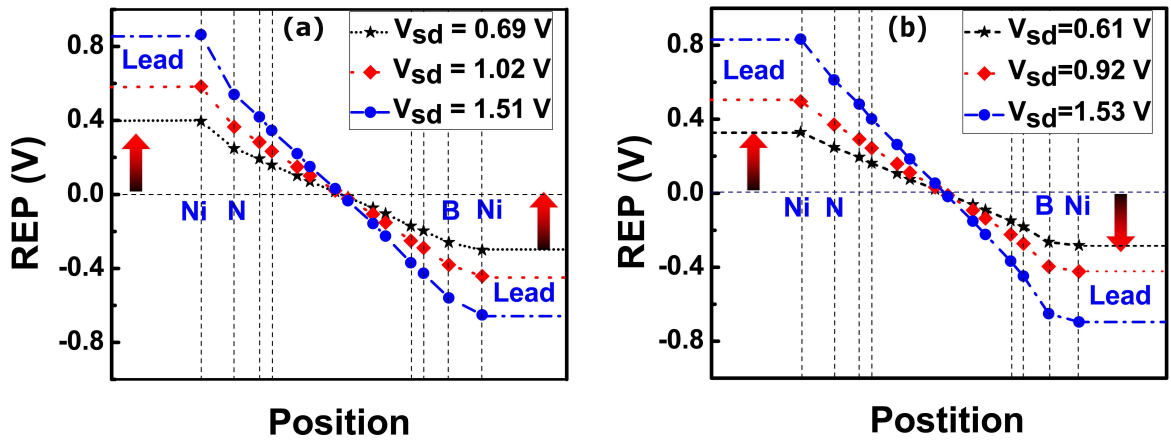, width=440pt}
\caption{ {\bf Bias dependent planar average  electrostatic potential profile  in BNTQD-Ni junction at different atomic positions.} 
(a) for PC and  (b) for APC. The vertical dotted lines represent the planar atomic position of the device along the direction of current;
the horizontal dotted line refers to the equilibrium situation; REP refers to the potential drop with respect to the equilibrium.} 
\end{figure}
\clearpage
}

\afterpage{%
\begin{figure}
\epsfig{figure=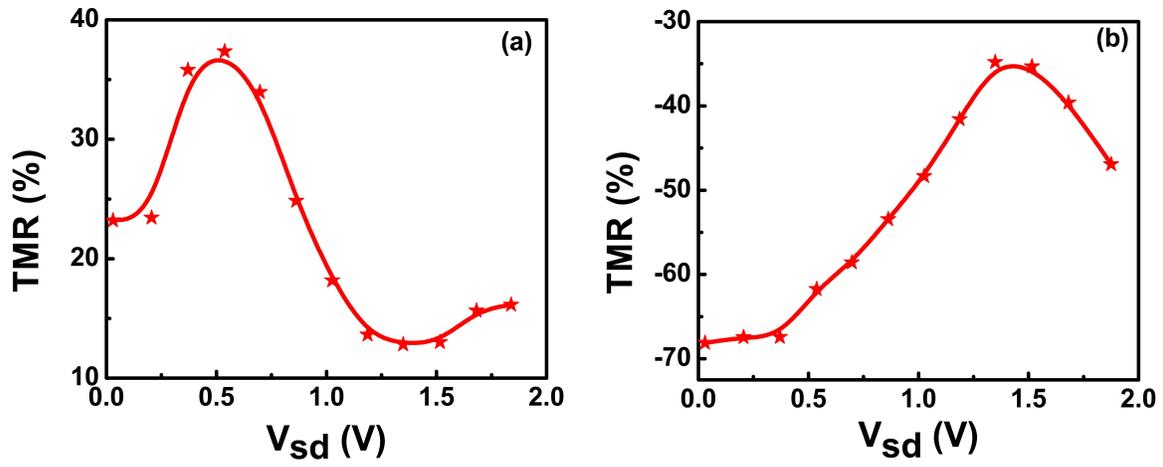, width=440pt}
\caption{{\bf Bias dependent tunnel magnetoresistance (TMR).} TMR as a function of applied bias ($V_{sd}$) for  
(a)  ${\varepsilon_g}$ = 0.00 V/\AA$ $, and  (b)  ${\varepsilon_g}$ = 2.04 V/\AA.}
\end{figure}
\clearpage
}

\afterpage{%
\begin{figure}
\epsfig{figure=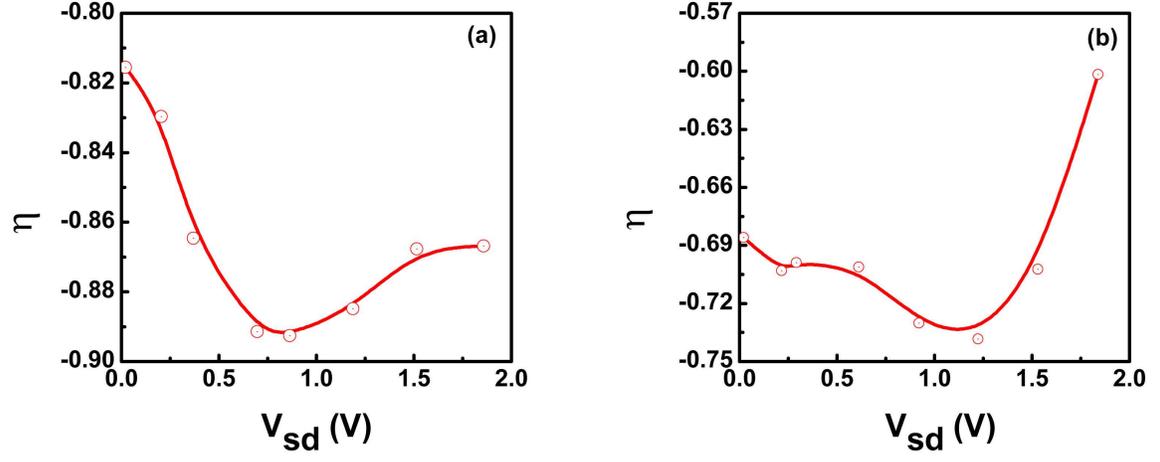, width=440pt}
\caption{{\bf Bias dependent spin injection factor ( $\eta$) in the BNTQD-Ni tunnel junction for $\varepsilon_g$ = 0 V/\AA.}  (a) parallel spin configuration (PC),
 (b) anti-parallel spin configuration (APC). Since  $I_{down}$ $>$ $I_{up}$ , the 
$\eta$ is found to be negative.}
\end{figure}
\clearpage
}
\afterpage{%

\begin{figure}
\epsfig{figure=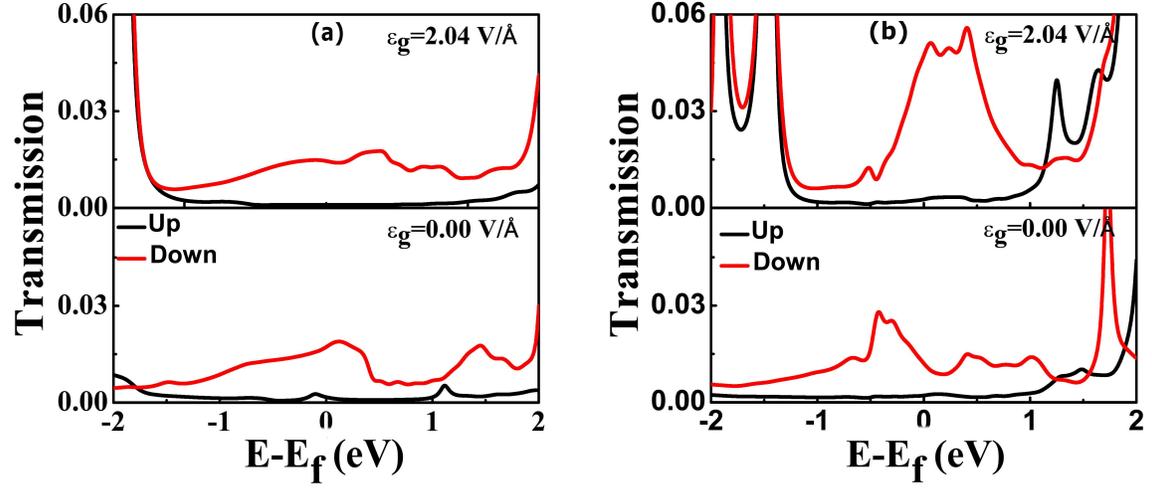, width=440pt}
\caption{ {\bf  Transmission function. } ${\varepsilon_g}$ dependent spin-polarized transmission  for (a) parallel spin configuration and (b) anti-parallel spin configurations at $V_{sd}$ of $\sim$ 0.2 V. }

\end{figure}
 \clearpage
 }
 
 \afterpage{%
\begin{figure}
\epsfig{figure=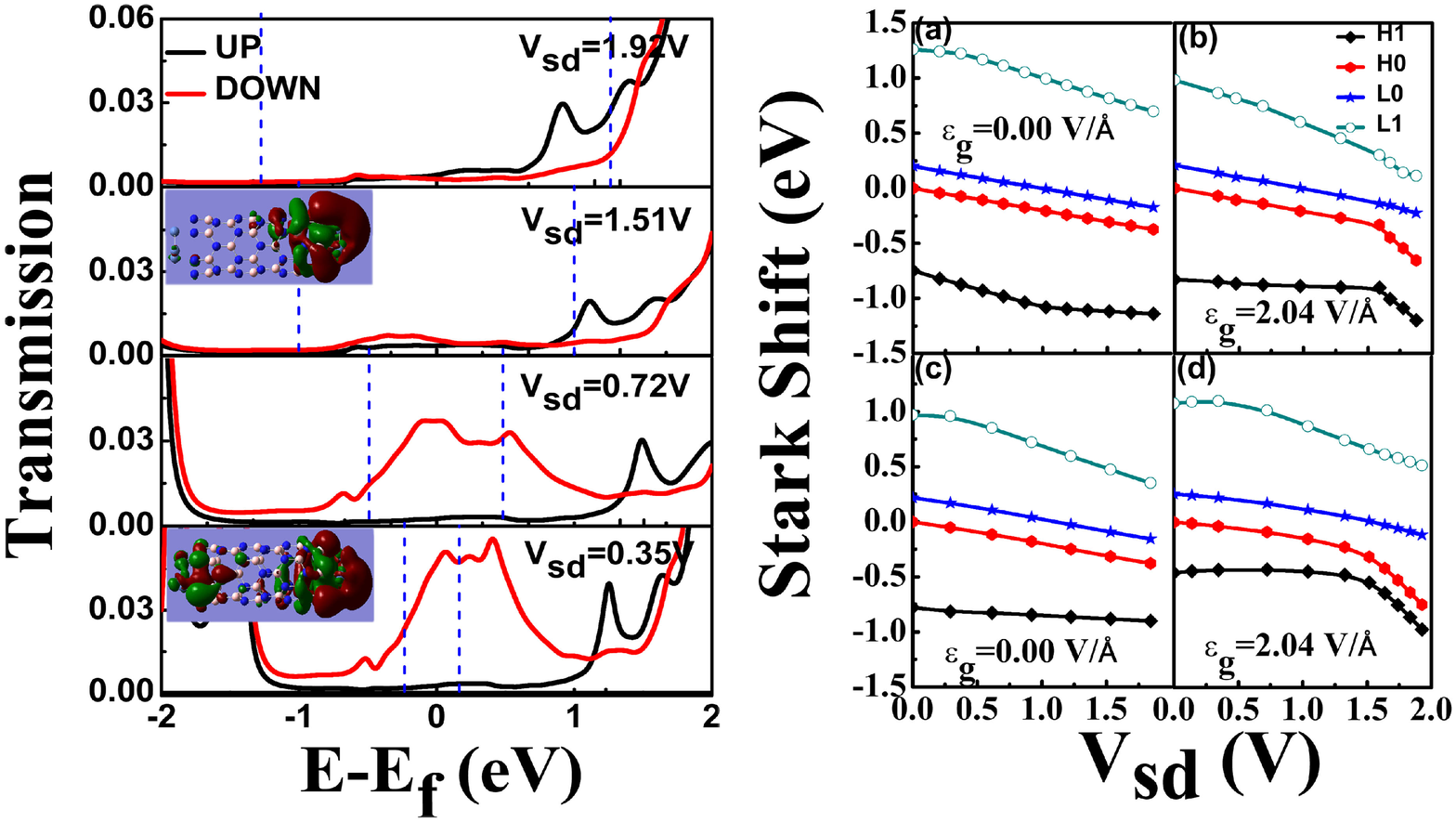, width=440pt}
\caption{{\bf Stark effect.}(a) Bias dependent spin-polarized transmission for the APC of the extended system
 at $\varepsilon_g$ = 2.04 V/\AA. Dotted lines
represent the  chemical potential window (CPW). The inset shows the bias dependent spin-down HOMO for the APC at $\varepsilon_g$ = 2.04 V/\AA. (b) Bias dependent
Stark shift  corresponding to the  frontier spin-down orbitals as a function of ${\varepsilon_g}$. 
Upper two panels are for PC and lower two panels are for APC.
H0, H1 refer to  the HOMO and HOMO-1, and  L0, L1 refer to the LUMO and LUMO+1  spin-down orbitals in the extended system.}
\end{figure}
\clearpage
}


\begin{thebibliography}{99}
\bibitem{Fert} A. Fert,   {\it Rev. Mod. Phys.},  2008, {\bf  80}, 1517-1530.
\bibitem{Ouyang} M. Ouyang and D. D.  Awschalom,  {\it Science}, 2003, {\bf 301},
1074-1078. 
\bibitem{Schmaus} S. Schmaus et al.,   {\it Nat. Nanotechnol.},  2001, {\bf 6},  185-189.

\bibitem{Hauptmann} J. R. Hauptmann, J. Paaske and P. E.  Lindelof,  {\it  Nature. Phys.},  2008, {\bf 4},  373-376.

\bibitem{Sahoo} S. Sahoo {\it et al.},   {\it Nature. Phys.}, 2005, {\bf 1},  99-102.

\bibitem{Yamada}  R. Yamada, M.  Noguchi and H. Tada,  {\it Appl. Phys. Lett.},  2011, {\bf 98}, 053110.

\bibitem{Horiguchi} K.  Horiguchi, T.  Sagisaka, S.  Kurokawa and A.  Sakaia,  {\it J.  Appl. Phys.},  2013, {\bf 113}, 144313-7.

\bibitem{Urdampilleta}  M. Urdampilleta, S. Klyatskaya, J-P Cleuziou and  M.   Ruben,  {\it Nature  Mater.}, 2011, {\bf 10},   502-506.

\bibitem{Barraud}  C. Barraud  {\it et al.},  {\it Nature Phys.},  2010, {\bf 6}, 615-620.

\bibitem{Baibich} M. N. Baibich  {\it et al.},  {\it Phys. Rev. Lett.},  1988, {\bf 61},  2472-2475.

\bibitem{Binasch} G. Binasch, P. Grunberg,  F. Saurenbach and  W. Zinn,  {\it Phys. Rev. B},  1989, {\bf 39},  4828-4830.

\bibitem{Parkin} S. S. P. Parkin, N. More and K. P. Roche,  {\it Phys. Rev. Lett.},  1990, {\bf 64},  2304-2307.

\bibitem{Dediu} V. A. Dediu, L.  Hueso, I.  Bergenti and  C. Taliani,  {\it Nature Mat.}, 2009, {\bf 8}, 707-716.

\bibitem{Kim} W. Y. Kim and  K.S. Kim,  {\it Acc.  Chem. Res.}, 2010, {\bf 43},  111-120.

\bibitem{Sanvito1}  S. Sanvito,  {\it  Chem. Soc. Rev.},   2011, {\bf 40},  3336-3355.
  
\bibitem{Liu}  D, Liu, Y.  Hu, H.  Guo and X. F. Han,  {\it  Phys. Rev. B},   2008, {\bf 78},  193307-193310.
 
\bibitem{Atodiresei}  N. Atodiresei {\it et al.},  {\it Phys. Rev. Lett.},  2010, {\bf 105},  066601-4.

\bibitem{Waldron}  D. Waldron, P. Haney, B.   Larade, A.  MacDonald and H. Guo,  {\it Phys. Rev. Lett.},   2006, {\bf 96},  166804-4.

\bibitem{Sanvito2} S. Sanvito,  {\it Nature Phys.}, 2010, {\bf 6}, 562-564.

\bibitem{Ruden} P. Ruden, {\it Nat. Mater.}, 2011, {\bf 10},  8-9.

\bibitem{Mandal}  S. Mandal and  R.  Pati,   {\it  ACS Nano.},  2012, {\bf 6},   3580-3588.


\bibitem{Tsukagoshi} K. Tsukagoshi, B. W.  Alphenaar and H.  Ago,  {\it Nature(London)},  1999, {\bf 401}, 572-574.

\bibitem{Wang}  B. Wang, Y. Zhu, W.  Ren, J. Wang and  H. Guo,  {\it  Phys. Rev. B},    2007, {\bf 75}, 235415.


\bibitem{Chopra}  N. G. Chopra  {\it et al.},  {\it  Science}, 1995, {\bf 269},  966-967.

\bibitem{Bando}  D. Golberg {\it et al.},   {\it ACS Nano.}, 2010, {\bf 4}, 2979-2993.


\bibitem{Ishigami} M.  Ishigami, J. D. Sau, S.  Aloni, M. L.  Cohen and  A.   Zettl, {\it  Phys. Rev. Lett.},  2005, {\bf 94},  056804.

\bibitem{Taylor} J. Taylor, H. Guo and  J. Wang,  {\it Phys. Rev. B}, 2001, {\bf 63}, 245407-13.

\bibitem{Lang} M. Di Ventra,  S. T. Pantelides, and N.D. Lang,  {\it Phys. Rev. Lett.},   2000,  {\bf 84},  979Ð982.

\bibitem{Brandbyge} M. Brandbyge, J. L. Mozos, P.  Ordejon, J. Taylor and K. Stokbro, {\it Phys. Rev. B}, 2002, {\bf 65}, 165401-17.

\bibitem{Xue} Y. Xue, S.  Datta and M. A.  Ratner,  {\it J. Chem. Phys.},  2001, {\bf 115},  4292-4299.

\bibitem{Su} W. Su, J. Jiang, W.  Lu and Y.  Luo,  {\it Nano Lett.},  2006, {\bf 6},  2091-2094.

\bibitem{Solomon} G. C. Solomon, C. Herrmann, T.  Hansen, V.  Mujica and  M. A. Ratner, {\it Nat. Chem.},  2010, {\bf 2},  223-228.

\bibitem{Gaussian}  GAUSSIAN09, revision A.1, Gaussian, Inc., Wallingford, CT, 2009. 

\bibitem{Sai} N. Sai, M. Zwolak, G. Vignale, M. Di Ventra, {\it Phys. Rev. Lett.},  2005, {\bf 94},  186810-4.

\bibitem{Runge} E. Runge and E. K. U. Gross,  {\it Phys. Rev. Lett.}, 1984, {\bf 52},   997-1000.

\bibitem{Song} H. Song  {\it et al.},  {\it Nature},  2009, {\bf 462},
1039-1043.

\bibitem{Datta} S. Datta,  {\it Electron Transport in Mesoscopic Systems} ( Cambridge
University Press: Cambridge, UK, 1997).

\bibitem{Ventra}  M. Di Ventra,  {\it Electrical Transport in Nanoscale Systems}
(Cambridge: New York, 2008).
\bibitem{Herrmann} C. Herrmann, G. C.  Solomon and M. A.  Ratner,  {\it J. Am. Chem. Soc.}, 2010, {\bf 132},  3682-3684.

\bibitem{Pati} R. Pati, L.  Senapati, P. M.  Ajayan and S. K.   Nayak,  {\it Phys. Rev. B},  2003, {\bf 68},  1004071-4(R).


\bibitem{Rocha}  A. R. Rocha {\it et al.}, {\it Nature Mater.}, 2005,  {\bf 4},  335-339.

\bibitem{Ning} Z. Ning, Y.  Zhu, J.  Wang and  H.  Guo, {\it  Phys. Rev. Lett.},  2008, {\bf 100},   056803-4.

\bibitem{Zwolak}  M. Zwolak and M. Di Ventra, {\it App. Phys. Lett.},  2002, {\bf 81}, 925-927.

\bibitem{Dalgleish}  H. Dalgleish and  G. Kirczenow, {\it Phys. Rev. B}, 2005, {\bf 72}, 184407-5.

\bibitem{Khoo}  K. H. Khoo, M. S. C.  Mozzoni and  S. G. Louie,  {\it Phys. Rev. B},  2004,  {\bf 69}, 201401(R).

\bibitem{Chen}  C-W. Chen, M-H  Lee and S. J.  Clark, { \it Nanotechnology},  2004, {\bf 15}, 1837-1843.


\bibitem{Kong} J. Kong {\it et al.},  {\it Phys. Rev. Lett.},  2001, {\bf 87}, 106801-4.

\bibitem{Xu} B. Xu,  X.  Xiao,  X. Yang, L.  Zang and  N. Tao,  {\it J. Am. Chem. Soc.}, 2004, {\bf 127},  2386-2387. 


\bibitem{Cinchetti} M.  Cinchetti {\it et al.},  {\it Natur Mater.},  2009, {\bf 8},  115-119.

\bibitem{McClain}  R. Pati, M.  McClain and A.  Bandyopadhyay, {\it Phys. Rev. Lett.},  2008, {\bf 100},  246801-4.

\end{thebibliography}
\end{document}